\documentstyle[pra,aps,preprint,epsf]{revtex}

\begin{document}
\title{Distant clock synchronization using entangled photon pairs}
\author{Alejandra Valencia, Giuliano Scarcelli and Yanhua
Shih}
\address{Department of Physics, University of Maryland, Baltimore
County, \\ Baltimore, Maryland 21250} \maketitle\date{14 january,
2004}

\begin{abstract}
We report a proof-of-principle experiment on distant clock synchronization.
Besides the achievement of picosecond resolution at 3 kilometer distance,
this experiment demonstrated a novel concept for high accuracy non-local
timing and positioning based on the quantum feature of entangled states.
\end{abstract}
\vspace{2cm}

Accurate timing and positioning metrological measurements are
important for both fundamental research and practical
applications. In particular, distant clock synchronization has
attracted a great deal of attention in recent years due to its
essential role in the Global Positioning System (GPS) and
telecommunications \cite{bahderbook}.

Modern clocks have been improved to such a level
\cite{scientamer}, that the resolution and accuracy of the
comparison techniques have become the limiting factors to
determine their relative rates and synchronization. There are two
standard methods for synchronizing two distant clocks: the classic
Einstein protocol \cite{Einsteinprotoc} and the Eddington slow
transportation method \cite{Eddingtonprotoc}.  Both methods have
certain limitations and difficulties in high accuracy nonlocal
synchronization in which relativistic effects, such as the
rotating disk problem, have to be taken into consideration.
Einstein protocol is a two-way method, hence, it requires (1) an
accurate knowledge of the one-way speed of light, that until now
has not been measured conclusively on rotating reference systems
and (2) the light propagation path to be the same in each
direction.  The Eddington transportation method relies on the
physical movement of a clock, therefore, this method is not
practical for space applications.  Recently, the nonlocal
characteristics of entangled states have brought attention for
possible protocols of high resolution clock synchronization
\cite{dowling} \cite{giovannature} \cite{giovannettipresent}.

In this letter we wish to report an experimental
proof-of-principle demonstration of a new concept, based on the
nonlocal feature of entangled states, that can be practically
implemented for certain metrology applications such as
high-accuracy one-way synchronization of clocks and high-accuracy
positioning.

Our method relies on the measurement of the second order
correlation function of entangled states. In particular, we
consider the entangled photon pairs produced in a continuous wave
(CW) pumped spontaneous parametric down conversion (SPDC)
\cite{DNK}. Very roughly speaking, the process of SPDC involves
sending a pump laser beam into a nonlinear material, such as a
non-centrosymmetric crystal. Occasionally, the nonlinear
interaction inside the crystal leads to the annihilation of a high
frequency pump photon and the creation of two lower frequency
photons named as signal and idler.   The creation time of either
signal photon or idler photon is unknown; however, if the signal
photon is registered at a certain time, the detection time of the
idler photon can only happen at an unique precise time.   In the
reported experiment, both the signal photon and the idler photon
are in the form of continuous wave, i.e., $\Delta t =\infty$,
nevertheless the time correlation measurement of the signal-idler
at distance of $3km$ has shown uncertainty in the order of
picosecond.

According to quantum
field theory, the probability of having a joint photo-detection
event at space-time points $({\bf r}_{1}, t_{1})$ and $({\bf
r}_{2}, t_{2})$ is proportional to the second-order correlation
function of the fields \cite{Glauber}
\begin{eqnarray}\label{G2-0}
G^{(2)}({\bf r}_{1}, t_{1};  {\bf r}_{2}, t_{2}) = \langle \,
E^{(-)}({\bf r}_{1}, t_{1}) E^{(-)}({\bf r}_{2}, t_{2})
E^{(+)}({\bf r}_{2}, t_{2}) E^{(+)}({\bf r}_{1}, t_{1}) \, \rangle
\end{eqnarray}
where $E^{(-)}$ and $E^{(+)}$ are the negative-frequency and the
positive-frequency field operators of the detection events at
space-time points $({\bf r}_{1}, t_{1})$ and $({\bf r}_{2},
t_{2})$. For the two-photon entangled state of SPDC, $G^{(2)}({\bf
r}_{1}, t_{1}; {\bf r}_{2}, t_{2}) $ can be written as the modulus
square of a two-photon effective wavefunction, or biphoton:
\begin{eqnarray}\label{square}
G^{(2)}({\bf r}_{1}, t_{1};  {\bf r}_{2}, t_{2}) &=&  | \left< \,0
\, \right|E^{(+)}({\bf r}_{2}, t_{2}) \, E^{(+)}({\bf r}_{1},
t_{1}) \left| \, \Psi \, \right>  |^{2}  \equiv   | \, \Psi({\bf
r}_{1}, t_{1};  {\bf r}_{2}, t_{2}) \, |^{2}
\end{eqnarray}
where $\mid 0 \, \rangle$ stands for the vacuum, and $\mid \Psi \,
\rangle$ is the state of the signal-idler photon pair
\cite{rubin}.

 The two-photon effective wavefunction is
calculated to be:
\begin{eqnarray}\label{2wavefunction-SPDC}
\Psi (r_{1}, t_{1}; r_{2}, t_{2}) = e^{-i (\omega_{s}^{0} \tau_{1} + \omega_{i}^{0} \tau_{2}) } \, {\mathcal F}_{\tau_{1}-\tau_{2}} \, \big\{ f(\Omega)\big\}
\end{eqnarray}
where ${\mathcal F}_{\tau_{1}-\tau_{2}} \, \big\{ f(\Omega)\big\}$
is the Fourier transform of the spectrum amplitude function
$f(\Omega)$, $\tau_{j}=t_{j}-r_{j}/u_{j}, j=1,2$, and $u_{j}$ is
the group velocity at frequencies $\omega_{s}^{0}$ and
$\omega_{i}^{0}$ along the optical paths 1 and 2, respectively.
$\omega^{0}_s$ and $\omega^{0}_i$ are the central frequencies of
the signal-idler radiation field.

The $G^{(2)}(r_{1}, t_{1}; r_{2}, t_{2})$ function for the
two-photon entangled state of SPDC is thus
\begin{equation}\label{G2-SPDC}
G^{(2)}(r_{1}, t_{1}; r_{2}, t_{2}) = | \, {\mathcal F}_{\tau_{1}-\tau_{2}} \, \big\{ f(\Omega)\big\} \, |^{2}.
\end{equation}
This function, depending on $\tau_{1}-\tau_{2}$ is independent of
the chosen reference coordinates:  it is a Lorentzian invariant
\cite{bahderrelativityofgps}. The spectrum amplitude function of
the SPDC, $f(\Omega)$, provides all the information about the
spectrum and the correlation properties of the signal-idler pair
and it has been well studied \cite{DNK} \cite{Aleja}. In the
collinear case, for type-II and nondegenerate type-I SPDC, the
spectral function is calculated as $f(\Omega) \sim
sinc(DL\Omega/2)$, where L is the length of the crystal and
$D=\frac{1}{u_{s}}-\frac{1}{u_{i}}$ is the inverse group velocity
difference for the signal and idler. For an $8mm$ LBO crystal
pumped at $458 nm$ (type-II),  the estimated width of
$G^{(2)}(t_{1}-t_{2})$ is about $800$ femtoseconds. For collinear
degenerate type-I SPDC, the spectral function is $f(\Omega) \sim
sinc (D^{''}L\Omega^{2}/2)$, where $D^{''}$ is the second
derivative of the dispersion function of the nonlinear material.
In this case, the width of $G^{(2)}(t_{1}- t_{2})$  is about $30$
femtoseconds for the same size LBO crystal. Typical values for the
natural width of $G^{(2)}$ for SPDC are, then, on the order of a
few femtoseconds to hundreds of femtoseconds. If $r_{1}$ and
$r_{2}$ are well controlled, the measurement of $t_{1}-t_{2}$ can
reach, in principle, the same order of resolution, making SPDC
particularly suitable for implementing protocols for timing and
positioning measurements with ultra high accuracy.

For example, consider a new protocol for one-way synchronization
of two distant clocks: we have clock-1 in a space station and
clock-2 in the laboratory (Fig.~\ref{Fig1}).  The signal and idler
photons are sent to two photon counting detectors $D_{1}$ (in
space) and $D_{2}$ (on the ground). The photon registration times
of the detectors, $t_1$ and $t_2$, are recorded by two ``event
timers" whose time bases are provided by clock-1 and clock-2,
respectively \cite{eventtimer}. The individual time history
records can be brought together through a classical communication
channel for comparison. If the two clocks are synchronized, the
joint detection of the signal-idler pair obtained by matching the
photon registration time records will show maximum
``coincidences".  If the clocks lose their synchronization, one
has to rematch the records to achieve maximum coincidences by
shifting one of them by a certain amount, that corresponds to how
much the two clocks have lost their synchronization. The clocks
can be adjusted and kept synchronized accordingly.

The initial
synchronization of the two clocks is made in the following way:
first we send the signal (with wavelength $\lambda_{s}$) to
detector $D_{1}$ and the idler (with wavelength $\lambda_{i}$) to
detector $D_{2}$. The registration time difference, $t_{1}-t_{2}$,
at $D_{1}$ and $D_{2}$ is estimated:
\begin{equation}\label{formulas1}
 t_{1}-t_{2}=\frac{r_{1}}{u_{s}}+t_{0}-\frac{r_{2}}{u_{i}}
\end{equation}
where $t_{0}$ is the time offset of the two non-synchronized clocks.
Then, we switch the signal and the idler, sending the idler to
$D_{1}$ and the signal to $D_{2}$. The registration time
difference, $t'_{1}-t'_{2}$ of $D_{1}$ and $D_{2}$ will now be
\begin{equation}\label{formulas2}
 t'_{1}-t'_{2}=\frac{r_{1}}{u_{i}}+t_{0}-\frac{r_{2}}{u_{s}}.
\end{equation}
Subtracting  the two registration time differences, we have
\begin{eqnarray}\label{formulas3}
\Delta t_{-}=(t_{1}-t_{2})-(t'_{1}-t'_{2}) =D \, (r_{1}+r_{2})
\end{eqnarray}
where $D=\frac{1}{u_{s}}-\frac{1}{u_{i}}$. $\Delta t_{-}$ is
obtained from direct measurements and we assumed that $r_{2}$ is
known ($D_{2}$ is in the laboratory). In some cases $D$ is known
or independently measurable, therefore the distance between the
space station and the laboratory, $r_{1}$, is predictable through
the measurements of $\Delta t_{-}$. In other cases, $r_{1}$ may be
given or independently measurable, so, the value of $D$ can be
calibrated in the above procedure with ultra-high accuracy. In
both cases substituting into either Eq.~(\ref{formulas1}) or
Eq.~(\ref{formulas2}), the time offset $t_{0}$ is thus estimated
with the same order of accuracy as the measurement of
$t_{1}-t_{2}$ and the clocks are synchronized accordingly. Notice
that the measurement can be easily repeated for different values
of $r_{2}$, so, for example, even in the case in which both $D$
and $r_{1}$ are unknown, measurements of $\Delta t_{-}$ with
different known values of $r_{2}$ allow the evaluation of $D$ and
$r_{1}$ simultaneously.

We performed the proof-of-principle experimental demonstration in
the case in which $r_{1}$ and $r_{2}$ were known: in the
laboratory, long optical fibers of known lengths were used to
simulate the nonlocal condition. The setup is shown in
Fig.~\ref{Fig2}. A single frequency $Ar^{+}$ laser line of $457.9
nm$ was used to pump an $8 mm$ LBO crystal for type-II SPDC. The
signal-idler radiations (centered at $\sim 901 nm$ and at $\sim
931 nm$ respectively) were separated from the pump laser beam by
using filtering devices. The orthogonally polarized signal-idler
pair was split by means of a polarizing beam splitter.  Before the
beam splitter, a half-wave plate was placed in order to perform
the two measurements described previously: when the waveplate is
at zero degrees, the signal is transmitted to $D_{1}$ and the
idler reflected to $D_{2}$; when the waveplate is at 45 degrees,
the idler is transmitted to $D_{1}$ and the signal is reflected to
$D_{2}$. In both measurements, the signal and idler radiation were
fed into two $1.5 km$ long commercial optical fibers optimized for
single-mode operation at $1300nm$. The signal-idler pair was then
detected by two single-photon counting detectors. After a large
number of signal-idler pair measurements, a histogram of the
number of counts against $t_{1}-t_{2}$ (the resolution of the
fast-timing electronics is $3 ps$) can be obtained. This
distribution function corresponds to the ${G}^{(2)}(t_{1}-t_{2})$
function previously described \cite{MW}.

Fig.~\ref{Fig3} shows the experimental results. The distribution
function on the left corresponds to the case of signal-$D_{1}$ and
idler-$D_{2}$ (half-wave plate at $0$ degree) while the
distribution function on the right corresponds to the case of
idler-$D_{1}$ and signal-$D_{2}$ (half-wave plate at $45$ degree).
The presence of multiple peaks on each individual distribution
function is a consequence of intermodal dispersion in optical
fibers, which is a known effect in fiber optics
\cite{fiberkeiser}.

The calculated width of the effective two-photon wavefunction from
a $8 mm$ type-II LBO SPDC, without the long optical fibers, is
about $800 fs$.  The measured width of ${G}^{(2)}(t_{1}-t_{2})$,
with the fibers, is around $750 ps$.  There are two contributions
for the broadening of the $G^{(2)}$ function: (1) dispersion in
the optical fiber, which may be compensated nonlocally
\cite{Franson} \cite{Aleja} (the compensation is not included in
this proof-of-principle experiment); (2) the time jitter of the
photo-detector.  The behavior of the biphoton in dispersive medium
has been previously studied \cite{Aleja}. Using two fibers of $1.5
km$ length, the far-field zone condition is satisfied. Therefore,
we expect $G^{(2)}(t_{1}-t_{2})$ to take the shape of the spectrum
function of the type-II SPDC $|\, f(\Omega)\, |^{2}$ with a FWHM
of $600 ps$. Fig.~\ref{Fig4} shows the central peak of the
experimental data (for the case of half-wave plate at 45 degrees)
compared with the theoretical expectation when the broadening
contributions of (1) and (2) are taken into consideration. The
fitting parameters $k''_{s}$ and $k''_{i}$  of the signal-idler
radiations, $2.76\times10^{-28}s^{2}/cm$ and
$2.96\times10^{-28}s^{2}/cm$, respectively, are in agreement with
the values specified by the manufacturer of the optical fiber.

By measuring the displacement of the central peak when the
half-wave plate is rotated from 0 to 45 degrees ( $\Delta
t_{-}=5432 \pm 1 ps$), and knowing the length of the fibers, the
experimental value for $D$, using Eq.~(\ref{formulas3}), was found
to be $1799.9\pm 0.4 ps/km$, in agreement with the parameters of
the fibers. Substituting the estimated value of $D$ into either
Eq.~(\ref{formulas1}) or Eq.~(\ref{formulas2}), the time offset is
measured to be $t_{0}=40369 \pm 1 ps$, which has the same order of
accuracy of the $t_{1}-t_{2}$ measurement.

In conclusion, we successfully performed a high accuracy (picosecond)
proof-of-principle experiment on distant clock synchronization ($3km$).
Besides the implementation of the novel one-way clock synchronization
protocol, this experiment has also demonstrated a novel concept of
quantum metrology for high accuracy nonlocal timing and positioning.

The authors thank M.H. Rubin, M. D'Angelo and D. Hudson for
helpful discussions; G. Carter and H. Jiao for the loan of the
fibers. This work was supported, in part, by NASA-CASPR program,
NSF, and ONR.

\begin{figure}[hbt]
\caption{Schematic setup of a novel protocol for one-way distant
clock synchronization.} \label{Fig1}
\end{figure}

\begin{figure}[hbt]
\caption{Schematic setup of the proof-of-principle experiment.}
\label{Fig2}
\end{figure}

\begin{figure}[hbt]
\caption{A typical histogram, the number of counts vs detection
time difference $t_{1}-t_{2}$.  Curve on the left: half-wave plate
at 0 degrees; Curve on the right: half-wave plate at 45 degrees.
The presence of side-peaks is due to the excitation of side-modes
in the fiber. } \label{Fig3}
\end{figure}

\begin{figure}[hbt]
\caption{Central peak of the measured distribution function of
$t_{1}-t_{2}$. The solid line is a theoretical fitting curve of
$G^{(2)}$ considering the broadening contributions of the
propagation dispersion and the jitter of the photon counting
detectors.} \label{Fig4}
\end{figure}

\centerline{\epsfxsize=2.5in \epsffile{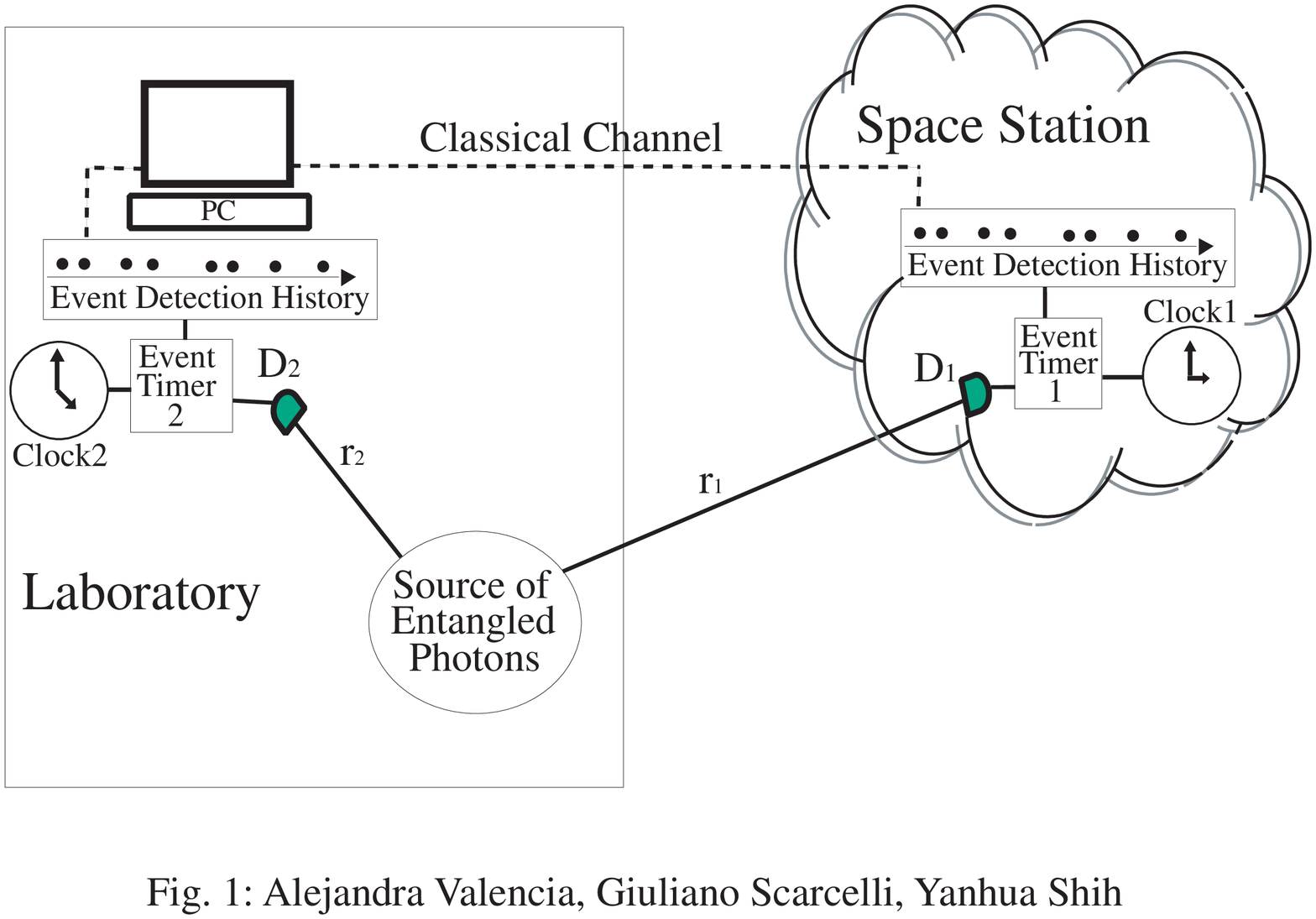}} \vspace{0.85cm}
Figure \ref{Fig1}. Alejandra Valencia, Giuliano Scarcelli, and
Yanhua Shih.

\centerline{\epsfxsize=3in \epsffile{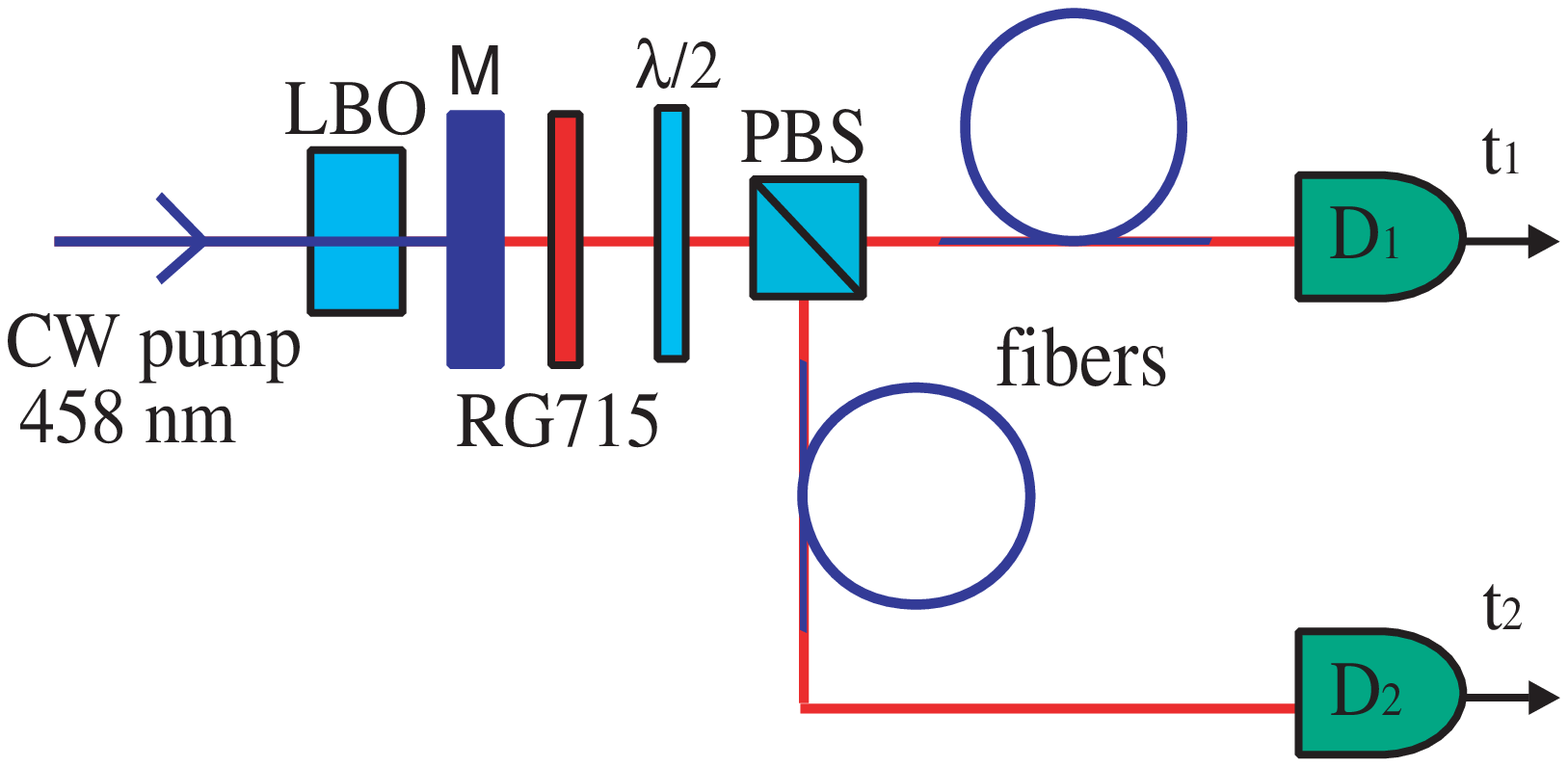}} \vspace{0.85cm}
Figure \ref{Fig2}. Alejandra Valencia, Giuliano Scarcelli, and
Yanhua Shih.

\centerline{\epsfxsize=2.8in \epsffile{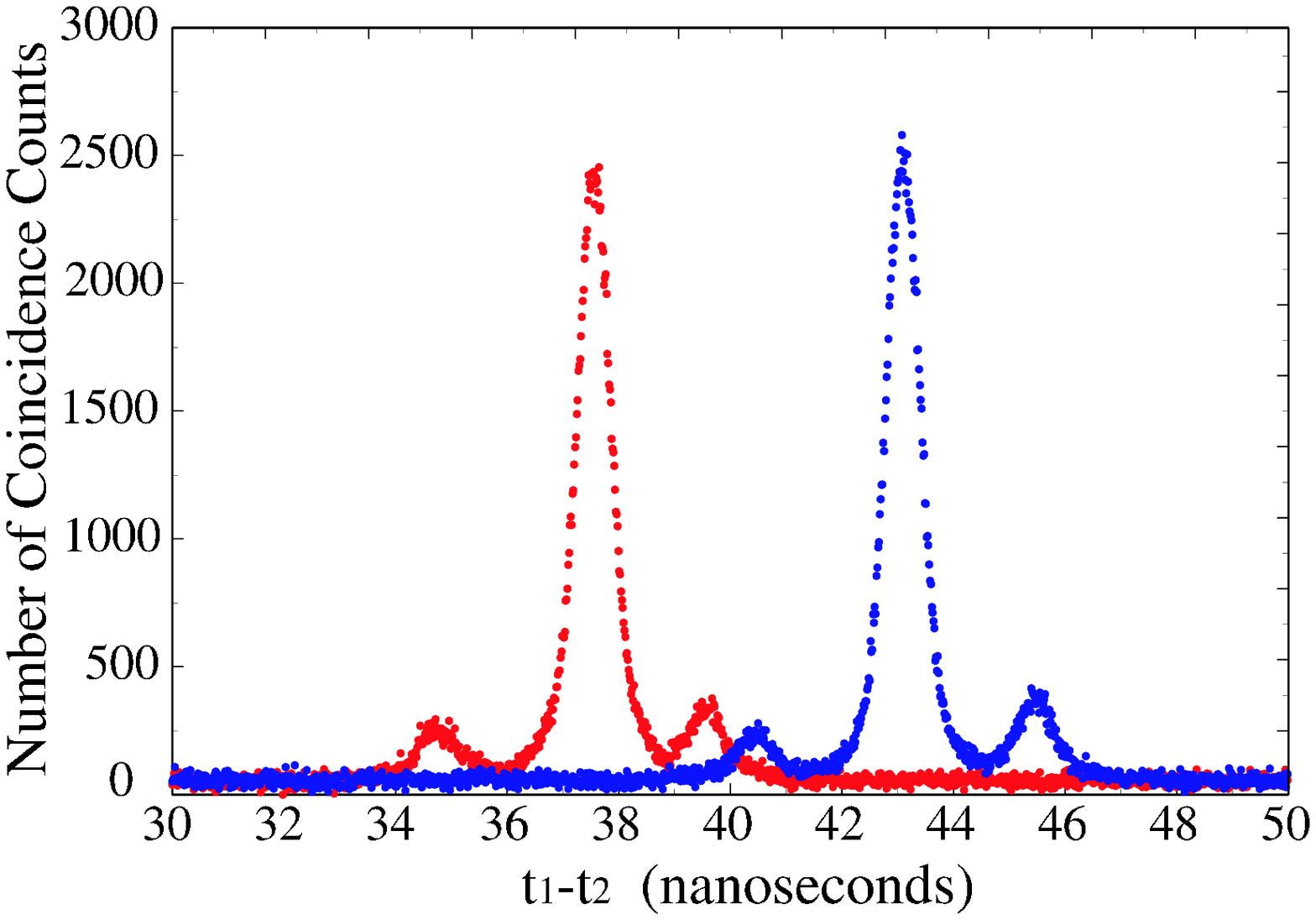}} \vspace{0.85cm}
Figure \ref{Fig3}.  Alejandra Valencia, Giuliano Scarcelli, and
Yanhua Shih.

\centerline{\epsfxsize=2.8in \epsffile{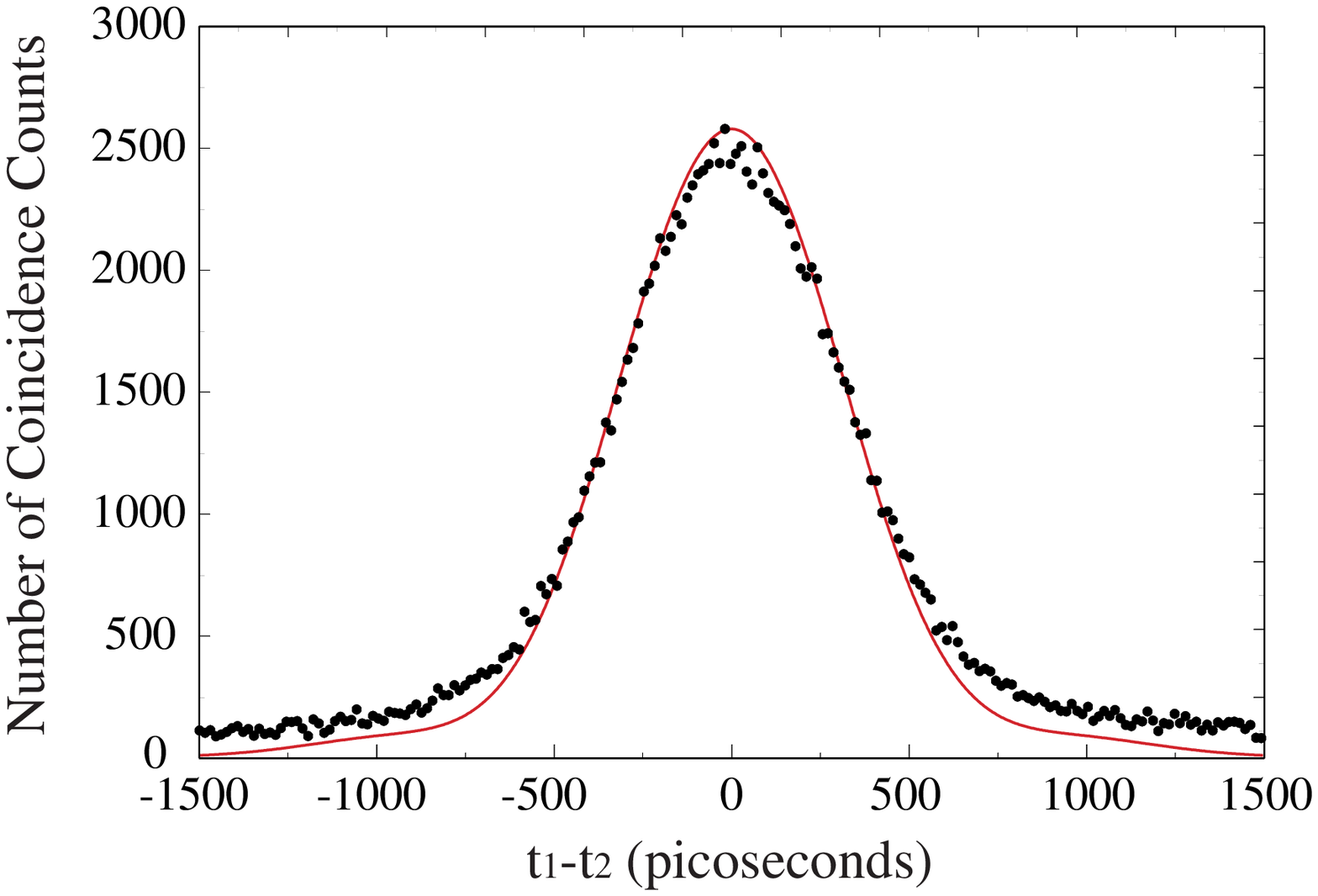}} \vspace{0.85cm}
Figure \ref{Fig4}.  Alejandra Valencia, Giuliano Scarcelli, and
Yanhua Shih.

\vspace{0.85cm}
\begin{references}

\bibitem{bahderbook} T. B. Bahder, "Clock Synchronization and
Navigation in the Vicinity of the Earth", gr-qc/0405001.

\bibitem{scientamer} W. Wayt Gibbs, Scientific American,  Special
issue, September 2002.

\bibitem{Einsteinprotoc} A. Einstein, Ann. Phys., {\bf 17}, 891
(1905).

\bibitem{Eddingtonprotoc} A. S. Eddington, "The Mathematical Theory of
Relativity", Chelsea Publishing Company, New York, (1924)

\bibitem{dowling} Richard Jozsa, Daniel S. Abrams, Jonathan P. Dowling, and Colin P. Williams.
Phys. Rev. Lett. {\bf 85}, 2010 (2000).

\bibitem{giovannature} Vittorio Giovannetti, Seth Lloyd and
Lorenzo Maccone, Nature. {\bf 412}, 417 (2001).

\bibitem{giovannettipresent} V. Giovannetti, S.Lloyd, L.Maccone,
and F.N.C.Wong, Phys. Rev. Lett. {\bf 87}, 117902 (2001).

\bibitem{DNK}D.N. Klyshko, {\em Photons and Nonlinear Optics}
(Gordon \& Breach, New York,1988).

\bibitem{Glauber} R.J. Glauber,
Phys. Rev. {\bf 130}, 2529 (1963); {\bf 131}, 2766 (1963).

\bibitem{rubin} M.H. Rubin, D.N. Klyshko, Y.H. Shih, and A.V. Sergienko, Phys. Rev. A {\bf
50}, 5122 (1994).

\bibitem{bahderrelativityofgps}Thomas B. Bahder, Phys. Rev.D {\bf
68}, 063005 (2003).

\bibitem{Aleja} A.Valencia, M.V.Chekhova, A.Trifonov and Y.Shih,
Phys. Rev. Lett. {\bf 88}, 183601 (2002).

\bibitem{eventtimer} C. Steggerda {\em et al}, Proceedings of the
10th International Workshop on Laser Ranging Instrumentation, Ed.
F.M. Yang, Chinese Academy of Sciences Press, 404 (1996).

\bibitem{MW} L.~Mandel and E.~Wolf, {\em Optical
Coherence and Quantum Optics}(Cambridge University Press, 1995).

\bibitem{Franson} J.D. Franson, Phys. Rev. A, 45, 3126 (1992).


\bibitem{fiberkeiser}  The fibers are optimized for single mode operation at $1300 nm$.
For the operating wavelengths (approximately 901 nm and 931 nm)
two modes are excited \cite{fibersenior}. These modes have different group velocities
that result in different time delays and causing the side peaks.

\bibitem{fibersenior} John M. Senior, {\em Optical Fiber Communications Principles and Practice}
(Prentice Hall International, 1992).

\end{references}
\end{document}